\documentclass[10pt,twoside]{hsqcd}
\usepackage{epsf,amsmath}
\usepackage{graphicx}

\begin{document}

\title{INTERFERENCE OF RESONANCES AND \\
OBSERVATION OF THE \boldmath$\Theta^+$-PENTAQUARK
}

\author{Ya.~AZIMOV\\ \\
 Petersburg Nuclear Physics Institute \\
Gatchina, Russia\\
E-mail:  azimov@thd.pnpi.spb.ru}

\maketitle

\begin{abstract}
\noindent
After brief discussion of how the quantum interference may
influence manifestations of resonances, description is
given of the recent experimental evidence for possible manifestation
of the $\Theta^+$-photoproduction in interference with the
$\phi$-photoproduction.

\end{abstract}

\markboth{\large \sl  Ya.~Azimov
\hspace*{2cm} HSQCD 2012} {\large \sl \hspace*{0.3cm}
Interference of resonances and
observation of the $\Theta^+$-pentaquark }

\section{General notes on quantum interference}

It is widely known that Quantum physics is probabilistic.
But this is not its most characteristic feature. Classical
physics may also be an object of probabilistic description
(such is, for example, statistical physics). The main difference
is that in cases where Classical physics adds probabilities,
Quantum physics adds amplitudes (or wave functions). Very
important consequences are existence of interference effects
and possible mixing of different wave functions (those are just
the most intimate properties of Quantum physics).

As an impressive result, some particles may oscillate in time
and space, transforming to each other. This quantum microscopic
effect may have quite macroscopic manifestations. Characteristic
oscillation distances can be as small as some mm's, or even
some $\mu$m's (for neutral $B$ mesons), but can also be
astronomically large (for solar neutrinos).

Hadron resonances can mix and oscillate as well, but their
space-time oscillations cannot be observed, since in all
realistic situations they are completely inside one atom
(all resonances have $c\tau<3\cdot 10^{-10}\,$cm; compare
to $c\tau=2.7\,$cm for $K_S$ mesons). Fortunately, the mixing
of resonances has visible manifestations in complementary
variables, first of all, in energy (or mass in the rest frame).
Here, mixing of resonances deforms their canonical Breit--Wigner
peaks.

\section{Manifestations of resonance interferences}

Typical examples of resonance interferences in various reactions
are collected and discussed in the topical review~\cite{az} (see
there references to the original works). This section briefly
follows to that review.

\subsection{Direct resonance interference}

There exist various cases and ways where and how two (or more)
resonances may interfere. The simplest (and best studied)
possibility, which may be called the direct interference,
appears when all decay products of one resonance can be produced
also in decays of another resonance. It may appear in different
reactions, but the situation most fruitful in this respect (and
easiest for description) arises in the $e^+e^-$
annihilation into hadrons, with different final states. It
demonstrates a great diversity of interference manifestations
(for details see, \textit{e.g.}, Ref.\,\cite{az}) .

Excitation curve of a resonance without any interference
corresponds to a peak which is traditionally described
by the Breit--Wigner formula. In the presence of interference,
familiar imagination for the excitation curve combines clear dip
and bump (in this or in the opposite order), corresponding to
destructive and constructive interferences. It is true indeed in
some cases, \textit{e.g.}, for the $\phi$ meson in the reaction
$e^+e^-\to\pi^+\pi^-\pi^0$ (interference with the $\omega$ meson,
see Fig.\,2 in Ref.\,\cite{az}).

However, it is not always so. Vicinity of the $\phi$ meson in the
reaction $e^+e^-\to\eta\gamma$ does not show a clear-cut dip.
Instead, the left and right foots of the $\phi$-peak are very
different (about ten times, see Figs. 3 and 4 in Ref.\,\cite{az}). This
is a result of interference with $\rho^0$ and $\omega$ mesons,
which makes the right foot of the $\phi$-peak lower than the left,
as a part of a broad dip.

Each of two above examples explicitly demonstrates both
constructive and destructive interferences. The pair of resonances
$\rho^0$ and $\omega$ (probably, the most famous example of
resonance interference) shows that only one kind of interference,
constructive or destructive, may be visible.

Indeed, in the reaction $e^+e^-\to\pi^+\pi^-$, where contribution
of the $\omega$ is very small, destructive interference ``bites
off'' part of the $\rho^0$-peak, without any visible $\omega$-peak
of constructive origin (see Figs. 5--7 in Ref.\,\cite{az}). On the
other side, in $e^+e^-\to\eta\gamma$, the $(\rho^0,\omega)$-peak
has its vertex at the $\omega$-mass, though the contribution of $\rho^0$
is larger. This is the result of constructive interference,
while destructive interference produces here only a barely visible break
in the left side of the combined peak (see Figs. 3 and 4 in Ref.\,\cite{az}).

Reaction $e^+e^-\to\pi^0\gamma$, where the $\rho^0$-contribution is
small in comparison with $\omega$, shows another picture. Here the
$\omega$-peak looks as an undistorted Breit--Wigner peak, but its
tails, both left and right, are enhanced by constructive interference
with the $\rho^0$-meson (see Fig.\,8 in Ref.\,\cite{az}). Similar is the
situation in the reaction $e^+e^-\to\pi^0\pi^+\pi^-$
(Fig.\,2~\cite{az}), where the $\rho^0$-contribution is also very
small, being suppressed by isospin violation.

In some cases interference can even transform a resonance peak
into dip. Such are manifestations of the $\phi$ meson in the reaction
$e^+e^-\to\omega\pi^0$, with subsequent different decays of the
$\omega$ (see Fig.\,9~\cite{az}).

Those examples demonstrate that the direct interference can distort
resonance peaks, sometimes very essentially. Effect for the same
resonance may appear different in different reactions and for different
decay modes. In any case, the direct interference became a good
instrument to search for (and study) rare decays of known resonances.

The reason is rather evident. If an amplitude is small, its direct
effect has the 2nd order smallness, while its interference has only
the 1st order smallness and may be additionally enhanced by multiplying
the small signal amplitude by a large background amplitude. Some rare
decay modes would never be discovered and measured without
interference. For example, the decay $\phi\to\omega\pi^0$ has
Br$\,\approx5\cdot10^{-5}$~\cite{kleo}, being twice
suppressed, by both the Zweig rule and isospin violation. Without
interference, it would be completely buried in fluctuations of
non-resonant events.

\subsection{Rearrangement interference}

Besides the direct interference, there is a possibility that only some
(or even one) of final particles may come from any of two interfering
resonances (it reminds the famous two-slit experiment, where a single
quantum particle may pass through one or another of two slits). Such kind
of interference can be called rearrangement (or rescattering) interference.

This phenomenon is known since 1960's. For example, processes
$~\pi^+p~\to~ \pi^+\Delta^+$,
$\pi^+p\to\pi^0\Delta^{++},~\pi^+p\to\rho^+p~$ produce, after decays of
resonances, the same final state $p\pi^+\pi^0$.  Therefore,
reaction $\pi^+p\to p\pi^+\pi^0$ should (and does) reveal specific
interference of the resonances $\Delta^{++},~\Delta^0,~\rho^+$ with
each other. Interference of such a kind may arise also in decays of
baryons or mesons with three or more hadrons in the final state.

Rearrangement interference, as well as direct one, distorts the resonance
peaks and spoils measurements of their masses and widths. To reject this
effect, kinematical regions, where the interference is most efficient, are
usually cut out. That is why the rearrangement interference continues to
be badly investigated and understood.

Nevertheless, it begins to be also used as an instrument for solving
various physical problems. It was applied, \textit{e.g.}, to eliminate
ambiguity in $CP$ violation studies of $B$ meson decays~\cite{babar}
(see also discussion in Ref.\,\cite{az}).

\section{Problem of the $\Theta^+$-pentaquark}

As a new step in similar direction, it was suggested to apply the
rearrangement interference to search for new resonances with small
production cross section~\cite{adp}. Specifically, the reaction
$\gamma p\to K^0\bar{K}^0p$ was suggested to look for the
signal of the $\Theta^+$ baryon in interference with the
$\phi$-photoproduction.

\subsection{Microreview: status of the $\Theta^+$}

After a number of discussions in framework of the Chiral Quark Model,
there appeared the first theoretical paper~\cite{dpp} which suggested
relatively certain properties for the strange baryon with $S=+1$, later
called $\Theta^+$: the mass $\sim1530~$MeV, decays to $pK^0$ and $nK^+$,
the total width $<15~$MeV. This baryon cannot consist of three quarks
as usual, it is the exotic pentaquark $uudd\bar{s}$.

Experimentally, there appeared about ten papers with evidence for the
$\Theta^+$, and about ten papers with negative results, some of them
having higher statistics (for references, see, \textit{e.g.},
Ref.\,\cite{am}). As a result, both Particle Data Group's position since
2008 and the common opinion of the high energy physics community are the
same: \textit{the pentaquark baryon is dead}\,! Strangely, this does not
prevent great enthusiasm in searches for exotic tetraquark mesons.

After 2008, some experimental collaborations withdrew their earlier
positive results, but others (LEPS and DIANA in particular) confirmed
observations of the $\Theta^+$ (see references in paper~\cite{am}).

Meanwhile, it was shown~\cite{ags} that all the data, both positive and
negative, can be reconciled, at least qualitatively, if multiquark (exotic)
hadrons are mainly produced from many-parton states (higher Fock components
of hadrons). Such states are always related to short-term fluctuations, and,
if this hypothesis is true, production of exotic hadrons may be considered
as a new kind of hard processes. Similar to all other hard processes, exotics
production should have small cross section.

Experimentally, smallness of the $\Theta^+$-production (if it exists at all)
was demonstrated by the CLAS analysis of the reaction $\gamma p\to K\bar
{K}p$~\cite{clas}. The $\Theta^+$ was not observed, and strict bound was
provided for its production cross section. This stimulated both the
suggestion~\cite{adp} and searches for an enhanced signal in rearrangement
interference, which resulted in the paper~\cite{am}.

\subsection{Evidence for a possible \boldmath$\Theta^+$-signal in
interference with the $\phi$ meson}

The new analysis of reaction $\gamma p\to K_SK_Lp$~\cite{am} used the
same data set as the earlier analysis~\cite{clas} and was, to some
extent, similar to it.  In both analyses one kaon was reconstructed by
the peak in the mass of $\pi^+\pi^-$ pairs, the other by the peak
in the missing mass $M_X(p\pi^+\pi^-)$.  But the analysis~\cite{am},
in difference with Ref.\,\cite{clas}, applied some additional
requirements to improve identification of the $K_S$. In both analyses
the $K_SK_L$ spectrum shows a very pronounced $\phi$-peak.
In Ref.\,\cite{clas} it was traditionally cut out, by applying the
condition $M_X(p)>1.04~$GeV. Analysis of
Ref.\,\cite{am}, just opposite, used events under the $\phi$-peak, with
$M_X(p)=1.02\pm0.01\,$GeV, where interference is most efficient.

The distribution in $M(pK_L)$, determined experimentally as $M_X(K_S)$,
shows now a peak when applying two additional cuts, separately or
together~\cite{am}. One of them restricts $M(pK_S)$ to eliminate known
$\Sigma^*\,$'s in the interval $1.5-1.7~$GeV, which otherwise provide strong
background in $M(pK_L)$ due to kinematical reflections. Another cut
restricts momentum transfers, to separate a definite (mainly
\begin{figure}[!thb]
\vspace*{0.3cm}
\centering
\includegraphics[width=0.7\hsize]{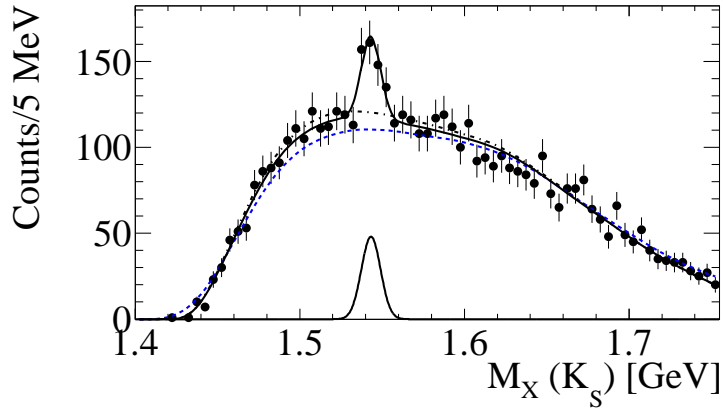}
\caption[*]{\small Distribution over $M_X(K_S)=M(pK_L)$ with cut
$|t_{\gamma K_S}|<0.45\,$GeV$^2$~\cite{am}}
\vspace*{-0.2cm}
\end{figure}
diffractive) mechanism of the $\phi$-production. An example of arising
spectra is shown in Fig.\,1. The background is described by the Monte
Carlo simulation based on the known Titov--Lee model for the forward
$\phi$-photoproduction off nucleon~\cite{tl}. The model is
theoretically meaningful (mainly Pomeron
exchange) and experimentally quite adequate. It well describes the whole
spectrum of Fig.\,1, except of a narrow peak near $M\sim1.54\,$GeV, its
width consistent with resolution. Statistical significance of the peak
is about 5.3$\,\sigma$. This peak should, of course, exist in $M(pK_S)$
as well, but it is not seen in this spectrum, because of worse
resolution (in agreement with the Monte Carlo simulation).

The strangeness of $pK_L$ is not fixed, it is either +1 or --1. Thus,
if the peak corresponds to a new baryon state, this baryon can be
either the $\Theta^+$ or a new $\Sigma^{*+}$. The latter case looks
less probable because of very small width and very low intensity of
production, unusual for $\Sigma^*\,$'s.  Moreover, $\Sigma^*$ should
also have hyperon decays without kaons.  However, no peak in $M_X(K_S)$
near 1.54\,GeV is seen when selecting the final state $K_SX$ without
$K_L$~\cite{am}. Both arguments prefer the $\Theta^+$, though do not
prove it. To finally prove that it is just the pentaquark baryon
$\Theta^+$, one needs to confirm existence of a direct photoproduction
signal (without interference) and then find the peak in a system with
the definite strangeness (\textit{e.g.}, in $nK^+$).

\section{Conclusions}

\begin{itemize}

\item Interference of resonances is a good instrument for solving
many problems.  Direct interference became familiar to search for and
study rare decays of known resonances. Rearrangement interference may
      be useful to amplify faint signals of known or unknown
      resonances, with any quantum numbers (the signal could be faint
      because of either small branching ratio or suppressed
      production).

\item A reliable signal has been found in rearrangement interference
      with the $\phi$ meson which may give evidence for possible
photoproduction of the pentaquark $\Theta^+$. The final confirmation
      awaits for finding both a direct signal of this process and a
      signal with definite strangeness. But even now one can rephrase
      Mark Twain's letter to say: ``The report of $\Theta^+\,$'s death
      was an exaggeration''.

\item Confirmation and investigation of multiquark hadrons may open new
      directions both for hadron spectroscopy and for QCD studies in general.
\end{itemize}

      \subsection*{Acknowledgements}
I thank my coauthors in Refs.\,[6] and [7], especially M.\,Amaryan and
 I.\,Strakovsky. This work was supported in part by the Russian
  State Grant No.\,RSGSS-65751.2010.2.

\end{document}